\documentclass[aps,pra,twocolumn]{revtex4-1}

\usepackage[normalem]{ulem}
\usepackage{amsmath,amssymb}
\usepackage{multirow}
\usepackage{xcolor,soul}
\usepackage{srcltx}
\usepackage{hyperref,graphicx}

\newcommand{\IFNPolimi}{Istituto di Fotonica e Nanotecnologie - Consiglio Nazionale delle Ricerche and\\ Dipartimento di Fisica - Politecnico di Milano,\\ Piazza Leonardo da Vinci, 32, I-20133 Milano, Italy}

\begin{document}

\title{Fractional Bloch oscillations in photonic lattices}

\author{Giacomo Corrielli}

\author{Andrea Crespi}

\author{Giuseppe Della Valle}

\author{Stefano Longhi}

\author{Roberto Osellame}
\email{roberto.osellame@ifn.cnr.it}

\address{\IFNPolimi}

\begin{abstract}
Bloch oscillations (BO), i.e., the oscillatory motion of a quantum particle in a periodic potential driven by a constant force, constitute one of the most striking and oldest predictions of coherent quantum transport in periodic lattices. In natural crystals,  BO have never been observed because of dephasing effects. Solely with the advent of semiconductor superlattices and ultracold atoms BO  have been observed for matter waves \cite{BOuper,BOultra1,BOultra2,BOultra3}. In their essence, BO are a wave phenomenon. As such, they are found for optical \cite{O1,O2,O3,O5,O6} and acoustic \cite{A1} waves as well. When interactions between particles compete with their mobility, novel dynamical behaviour can arise where particles form bound states \cite{B1} and co-tunnel through the lattice \cite{C1}. While particle interaction has been generally associated to BO damping \cite{K1,K2,K3}, for few strongly-interacting particles it was predicted that bound states undergo fractional BO at a frequency twice (or multiple) that of single-particle BO \cite{CB1,CB2,CB3}. The observation of fractional BO is challenging in condensed-matter systems and up to now has not been achieved even in model systems. Here we report on the first observation of fractional BO using a photonic lattice as a model system of a few-particle extended Bose-Hubbard Hamiltonian. These results pave the way to the visualization of other intriguing phenomena involving few correlated particles, such as the interplay between particle interaction and Anderson localization \cite{Flach,Steve1}, quantum billiards and transition to quantum chaos \cite{Kom11}, anyonic BO \cite{anyonic}, and correlated barrier tunneling and particle dissociation \cite{kolo2012}.
\end{abstract}

\maketitle

Ultracold quantum gases in optical lattices have provided in the past decade a powerful laboratory tool to simulate Hubbard models of condensed-matter physics  \cite{BlochReview}.  For few interacting particles, experiments with ultracold atoms have so far demonstrated the existence of bound particle states and correlated tunneling phenomena \cite{B1,C1}, however fractional BO of atomic pairs have not been observed yet. A photonic simulator of the Hubbard model can map the dynamics of two correlated particles hopping on a one dimensional lattice into the motion of a single particle in a two-dimensional lattice with engineered defects \cite{Longhi11,Kom11}, thereby offering the possibility to easily visualize the effect under simulation, but requiring a high level of control of the photonic structure.\par 

To study correlated BO, we consider an extended Bose-Hubbard (EBH) model \cite{E1,E3} describing strongly interacting bosons in the lowest Bloch band of a one dimensional lattice driven by an external force. This model is more accurate than the standard Bose-Hubbard (BH) model as it accounts for higher-order processes whose magnitude is comparable with the one of second-order tunneling \cite{E3}. The Hamiltonian of the system is given by
$\hat{H}=\hat{H}_{EBH}+\hat{H}_{F}$, where $\hat{H}_F  =   Fd  \sum_l l \hat{n}_{l}$
describes the effect of the external constant force $F$ ($d$ is the lattice period) and $\hat{H}_{EBH}=\hat{H}_{BH}+\hat{H}_3+\hat{H}_4+\hat{H}_5$ is the EBH Hamiltonian. $\hat{H}_{EBH}$ includes the standard BH Hamiltonian  (with $ \hbar=1$)
\begin{equation}
\hat{H}_{BH} = \frac{U_0}{2} \sum_l \hat{n}_l \left( \hat{n}_l-1 \right) -\frac{J}{2} \epsilon \sum_l \left( \hat{a}^{\dag}_l \hat{a}_{l+1}+ {\rm H.c.} \right)
\end{equation}
and three additional terms \cite{ E1}, whose explicit expression is given in the Supplementary information.
In Eq.(1), $U_0$ is the on-site energy interaction strength ($U_0>0$ for a repulsive interaction); $J$ is related to the single-particle hopping rate between adjacent lattice sites; $ \epsilon \ll 1$ is the lattice attenuation factor (see \cite{E1});  $\hat{a}_l^{\dag}$ and $\hat{a}_l$ are the bosonic creation and annihilation operators and $\hat{n}_l=\hat{a}^{\dag}_l \hat{a}_l$ is the particle number operator at lattice site $l$. The standard BH term $\hat{H}_{BH}$ describes on-site particle-particle interaction and single-particle tunneling between adjacent lattice sites [see Figs.1(a) and (b)].  The additional  terms $\hat{H}_3$, $\hat{H}_4$ and $\hat{H}_5$ account for nearest-neighbor hopping of single atoms conditioned
by the on-site occupation number, nearest-neighbor interactions, and nearest-neighbor
hopping of bosonic pairs, respectively [see Figs.1(c) and (d)].\par
\begin{figure*}
\includegraphics[width=14cm]{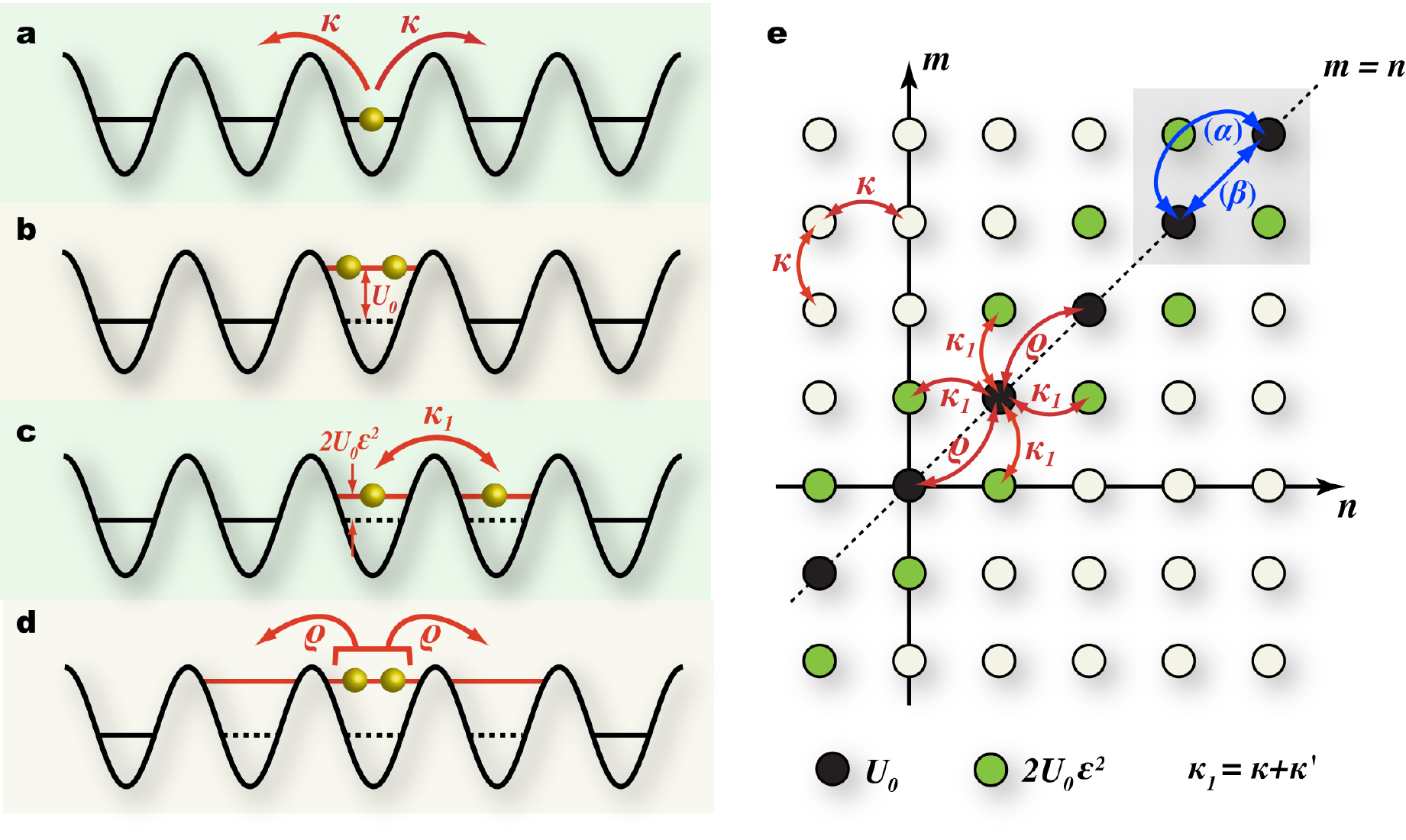}
\caption{{\bf Dynamics of two interacting bosons in the extended Bose-Hubbard lattice}.  Two-particle dynamics in the EBH model: (a) Single-particle tunneling; (b) on-site particle interaction; (c) nearest-neighbor-site particle interaction and conditional single-particle tunneling; (d) nearest-neighbor hopping of particle pairs. (e) Fock-space representation of the two-particle dynamics; the upper-right inset represents the two possible pathways for two-particle hopping to the nearest-neighbor site: ($\alpha$) is the second-order pair tunneling, while ($\beta$) is the direct two-particle tunneling.}
\end{figure*}

In the absence of particle interaction, each particle undergoes independent BO on the lattice. The single-particle amplitude probabilities $A_n(t)$ to find the particle at lattice site $n$ evolve according to the coupled equations
\begin{equation}
i \frac{d A_n}{dt}= -\kappa (A_{n+1}+A_{n-1})+FdnA_n
\end{equation}
where $\kappa= \epsilon J/2$ is the single-particle hopping rate. The single-particle energy spectrum is given by an equally-spaced Wannier-Stark ladder, which results in a periodic  dynamics at the BO frequency $\omega_B=Fd$. For single-site excitation, i.e. $A_n(0)=\delta_{n,0}$, BO appear as a breathing (rather than an oscillatory) motion at the frequency $\omega_B$.   Photonic demonstrations of such BO breathing dynamics were previously reported in Refs.\cite{O1,O2,O5}.\par 

Let us now consider the dynamics of two interacting particles. The state vector $| \psi(t) \rangle$ of the system can be expanded in Fock space as $| \psi(t) \rangle = (1 / \sqrt{2}) \sum_{n,m} c_{n,m}(t) \hat{a}^{\dag}_n \hat{a}^{\dag}_m | 0 \rangle$,
where $c_{n,m}(t)$ is the amplitude probability to find one particle at site $n$ and the other particle at site $m$, with $c_{n,m}=c_{m,n}$ for bosons.
The evolution equations for the amplitude probabilities (see the Supplementary information) show that the dynamics of the two correlated bosons on the one-dimensional lattice can be mapped into the motion of a {\it single} particle hopping on the two-dimensional square lattice of Fig.1(e). The square lattice shows both energy site defects and hopping rate corrections on the three diagonals $m=n,n \pm 1$. Far from the main diagonal, the square lattice is homogeneous with a uniform hopping rate $\kappa= \epsilon J/2$, which is the single-particle hopping rate of the original problem [see Fig.1(a)].  
\begin{figure*}
\includegraphics[width=14cm]{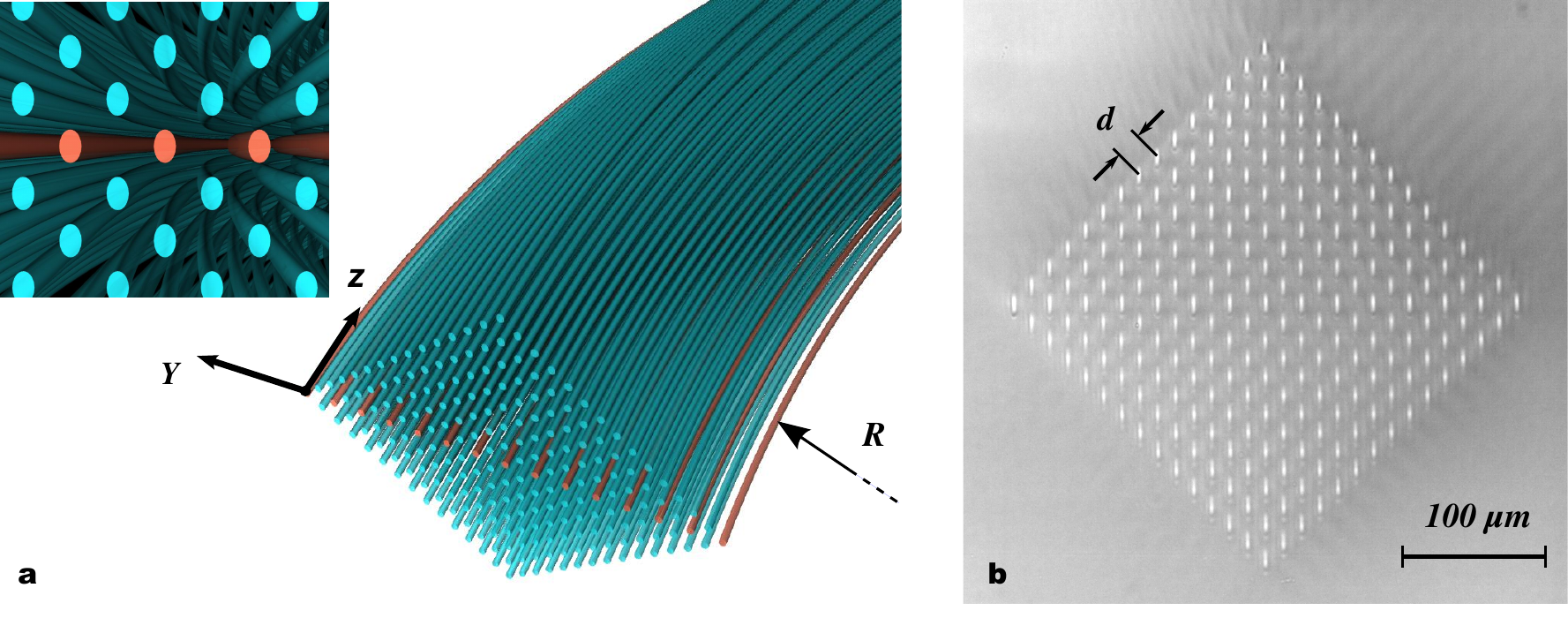}
\caption{{\bf The photonic simulator of correlated BO.} (a) Sketch of the waveguide array structure; red-coloured waveguides have a different refractive index change with respect to the others to implement the on-site particles-interaction defect $U_0$. (b) Section of the fabricated array, imaged with an optical microscope.}
\end{figure*}
The site energy defects $U_0$ and $ 2 \epsilon^2 U_0$ on the main ($m=n$) and nearest-neighbor ($m= n \pm 1$) diagonals account for on-site and nearest-neighbor-site particle repulsion, respectively [Figs.1(b) and (c)]. The correction of hopping rates $\kappa_1=\kappa+\kappa'$ (with $\kappa'=-U_0 \epsilon^{3/2}$) on the main diagonal accounts for nearest-neighbor hopping of single atoms conditioned by the on-site occupation number [Fig.1(c)]. Finally, the cross coupling $\rho=-2 U_0 \epsilon^2$ on the main diagonal $n=m$  describes
nearest-neighbor hopping of bosonic pairs, i.e. direct two-particle tunneling [Fig.1(d)].  For BO of two strongly-correlated particles ($ J \sim U_0$), only this last term is important, competing with second-order pair tunneling. In fact, let us assume as an initial condition $c_{n,m}(0)=A_n(0) \delta_{n,m}$, i.e. that the two particles are initially placed on the same site, with a probability $|A_n(0)|^2$ to find both particles at site $n$. Then an asymptotic analysis of the underlying equations shows that the two particles form a bound state, i.e. they co-tunnel along the lattice, undergoing BO at the frequency $ 2 \omega_B=2 Fd$ twice the single-particle BO frequency $\omega_B$. In Fock space, this means that the excitation remains confined along the main diagonal $m=n$ of the square lattice of Fig.1(e), and periodic revivals at frequency $2 \omega_B$ are expected. Precisely, as shown in the Supplementary information, the bound-particle occupation amplitudes $A_n(t) \sim c_{n,n}(t)$ at the various lattice sites evolve according to Eqs.(2), but with $F$ replaced by $2F$ and $\kappa$ replaced by $\kappa_{eff}$. The effective hopping rate of the bound particle state is given by
$k_{eff}=-2 \kappa^2/U_0+\rho=-\epsilon^2J^2/(2 U_0)-2 U_0 \epsilon^2$, which  includes  second-order tunneling ($-2\kappa^2 /U_0$) and direct two-boson tunneling ($ \rho$) processes. Such two contributions always add with the same sign, for both repulsive ($U_0>0$) and attractive ($U_0<0$)  interactions, and generally are of the same order of magnitude. The bound particle state thus behaves like a single particle hopping on the one-dimensional lattice, but with a driving force which is doubled  as compared to that of a single particle \cite{CB1,CB2} and with a modified hopping rate $\kappa_{eff}$.  Note  that the effect of direct two-atom tunneling on the BO dynamics, not considered in previous works \cite{CB2,CB3,Longhi11}, is to increase the amplitude of the breathing motion (owing to the correction of $\kappa_{eff}$), whereas the Wannier-Stark energy spectrum is not altered, i.e. $\omega_B$ is not affected. \par

In our experiments, single-particle hopping dynamics in the square lattice of Fig.1(e) has been simulated by discretized spatial light transport in an engineered two-dimensional square lattice of evanescently-coupled optical waveguides \cite{Lederer}. This lattice has been fabricated in a fused silica substrate by direct waveguide writing with femtosecond lasers \cite{OsellameSpringerBook, Gattass, CrespiPRL, SzameitJPB10, SzameitNP09}, taking advantage of the three-dimensional capabilities of this technology. The {\it spatial} propagation of the light intensity in the generic $(m,n)$-th lattice site maps the {\it temporal} evolution of the two-particle probability distribution $|c_{m,n}(t)|^2$ in Fock space \cite{Lederer,SzameitJPB10}. In the photonic simulator of $\hat{H}_{EBH}$, the role of the hopping coefficients $\kappa$ and $\rho$ is played by the evanescent coupling constants between first- and second-neighboring waveguides ($\bar{\kappa}$ and $\bar{\rho}$), respectively. The former is in general greater than the latter and the relative weight can be tuned by properly engineering the lattice parameter $d$ (see Supplementary Information). The on-site particle-interaction defect $U_0$ is realized by fabricating the main diagonal waveguides with a different refractive index change (this is achieved by slightly modifying the writing speed of the waveguides) and thus with the propagation constant $\beta'$ slightly detuned with respect to that of all the others ($\beta$). The positive or negative sign of the relative detuning $\Delta\beta = \beta-\beta'$ determines the repulsive or attractive nature of the interaction, respectively. Lastly, we did not add any special feature in the nearest-neighboring diagonals, because, as already discussed above, the nearest-site effects described in Fig.1(c) are negligible with respect to the other terms of the EBH model. The implementation of the driving force term $\hat{H}_{F}$ is obtained by a constant bending of the waveguide axes (radius of curvature $R$, $F\propto1/R$) in the plane determined by the lattice main diagonal direction $Y$ and the light propagation direction $z$ \cite{O5}.
In Fig.2(a) the whole lattice structure is depicted. Note that the array cross-section is rotated in such a way that the bending plane is horizontal in the laboratory reference frame [45$^{\rm o}$ rotation with respect to Fig.1(e)]. This simplifies the fabrication process and the imaging of the light propagating in the waveguides; in this way, although being a 3D structure, each waveguide remains on a specific horizontal plane.\par

In a first experiment we checked that our photonic simulator can reliably mimic the dynamics governed by $\hat{H}_{EBH}$ alone, namely without the action of the external constant force. For this purpose we fabricated 225 straight waveguides ($R=\infty$) arranged in a 15x15 square lattice, with spacing $d = 19 \mu m$ and a length $L = 2.5 cm$. For such a lattice spacing the coupling coefficients are $\bar{\kappa} = 0.95 cm^{-1}$ and $\bar{\rho} = 0.3 cm^{-1}$. To simulate particle interaction, we fabricated the main diagonal waveguides with a propagation constant detuning $\Delta \beta = - 4 cm^{-1}$, therefore representing an attractive interaction between the particles. A microscope image of the array section is reported in Fig.2(b). We coupled a $633 nm$ He-Ne laser radiation into the central waveguide, belonging to the main diagonal subset and thus corresponding to an initial condition where the two particles occupy the same lattice site. As previously discussed light is expected to remain on the detuned diagonal, representing the fact that the correlated particles will hop together to the next site and will never separate. To study the evolution of the system we  monitored the light propagation along the array by imaging from above the fluorescence light at ~650 nm emitted from the waveguides and arising from colour centres created in the femtosecond laser writing process. Since the colour centres are formed exclusively inside the waveguides, this technique yields a high signal-to-noise ratio and permits a direct imaging of the light propagation in the waveguide arrays \cite{SzameitJPB10, SzameitNP09}.
\begin{figure}
\includegraphics[width=8cm]{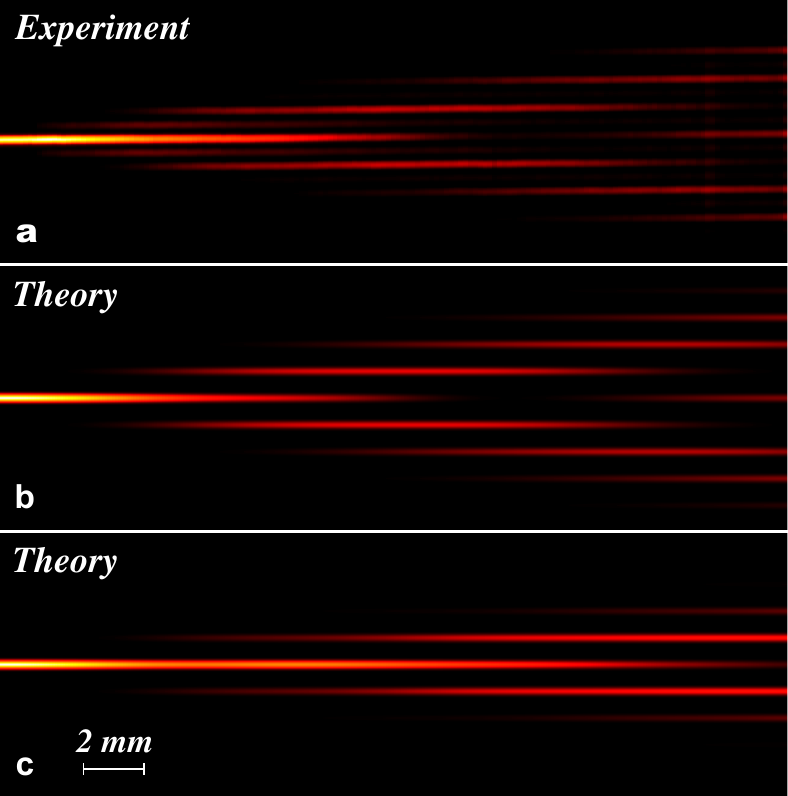}
\caption{{\bf Photonic simulation of the delocalization dynamics of two interacting particles in a crystal.} (a) Experimental simulation in a waveguide lattice implementing direct two-particle tunneling and second-order pair tunneling with a similar weight. (b) Numerical simulation including both contributions. (c) Numerical simulation including only second-order pair tunneling.}
\end{figure}
The result obtained in this first experiment is visible in Fig.3(a). The measured distribution of light in the photonic model implementing both contributions to the bound state hopping rate, i.e. two-particle tunneling and second-order pair tunneling, is compared with the corresponding numerical simulation [Fig.3(b)]. The excellent agreement demonstrates that our photonic model can indeed take into account both contributions and thus correctly implements the EBH model. The role of the direct two-particle tunneling [Fig.1(d)], neglected in the standard BH model, can be quite relevant as shown by the different light distributions obtained from numerical simulations of the same structure but including only the second-order tunneling [Fig.3(c)].\par
\begin{figure}
\includegraphics[width=8cm]{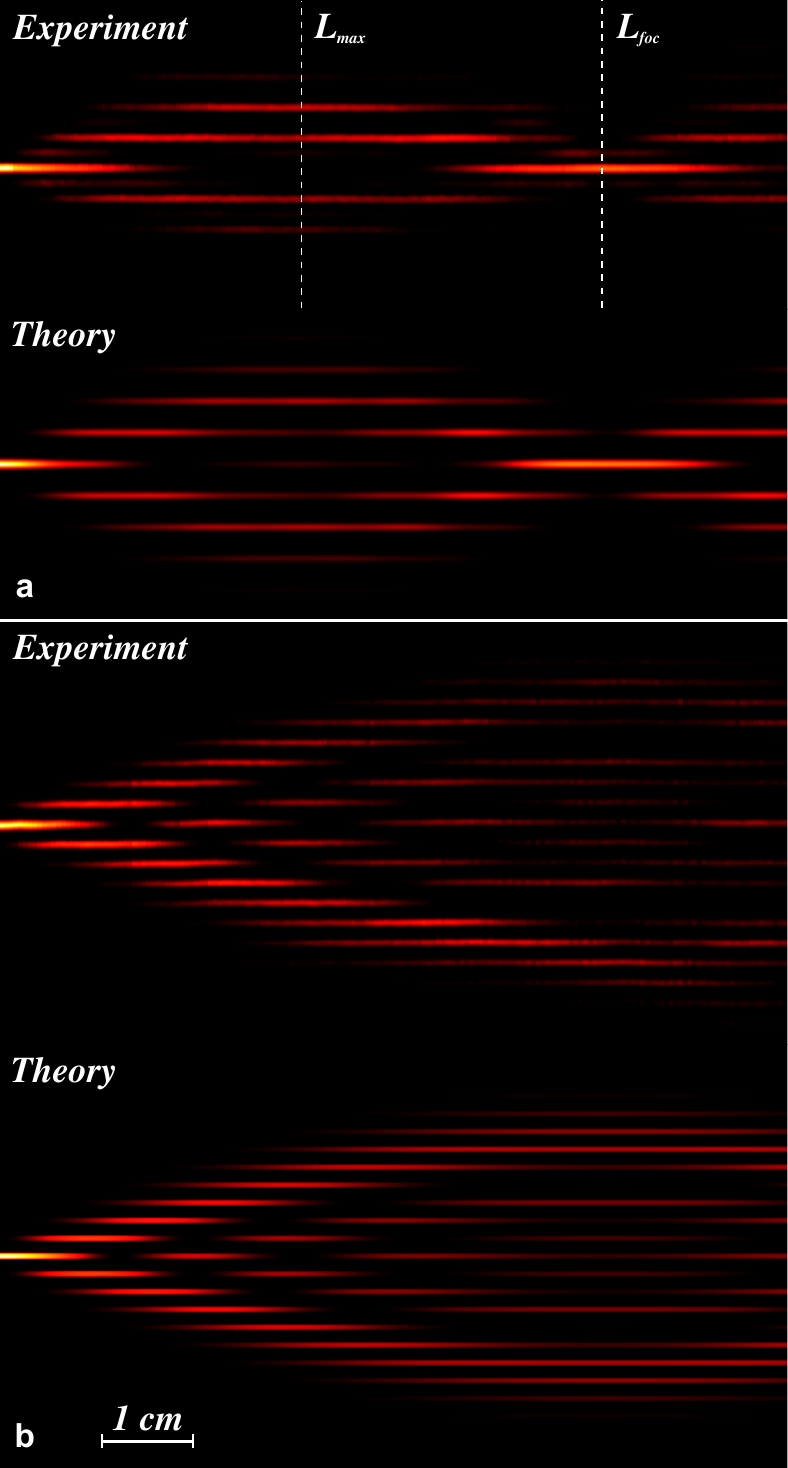}
\caption{{\bf Photonic simulation of Bloch oscillations for correlated and single particles.} a) Experimental and numerical simulation of the light distribution in the waveguide lattice diagonal, representing in Fock space the BO for two interacting particles . (b) Experimental and numerical simulation of the light distribution corresponding to single particle BO; the position where the BO amplitude is at maximum corresponds to the refocusing condition for the two-particle BO $L_{foc}=6.5 cm$.}
\end{figure}

In a second experiment, we proceeded to the observation of fractional BO. To this purpose we fabricated another 15x15 square waveguide lattice, with the same lattice spacing and detuning on the diagonal as in the previous structure. To mimic the external force $F$,  the waveguides are now circularly-bent with a radius $R=400 cm$, as depicted in Fig.2(a). The array length is $L=8.5 cm$. As in the previous experiment, we injected the probe light into the central waveguide and we imaged its propagation along the waveguide lattice diagonal (Y), where light is confined. Figure 4(a) shows the oscillatory behaviour of the light dynamics along the waveguide array.  The accuracy of the fabricated photonic simulator is confirmed by the very good agreement with the numerical simulations of light propagation in the designed structure [Fig.4(a)]. Initially, light spreads into several waveguides, until it reaches a maximum of the breathing amplitude (around $L_{max}=3.25 cm$). Then light refocuses into the central waveguide ($L_{foc}=6.5 cm$) and the breathing starts again. This periodic behavior represents the correlated BO of two interacting particles, watched in Fock space. To better analyze this result we cut the sample at $L_{max}$ and $L_{foc}$ and we imaged onto a CCD camera the light distribution exiting the output facet of the array (Fig.5).
\begin{figure}
\includegraphics[width=8cm]{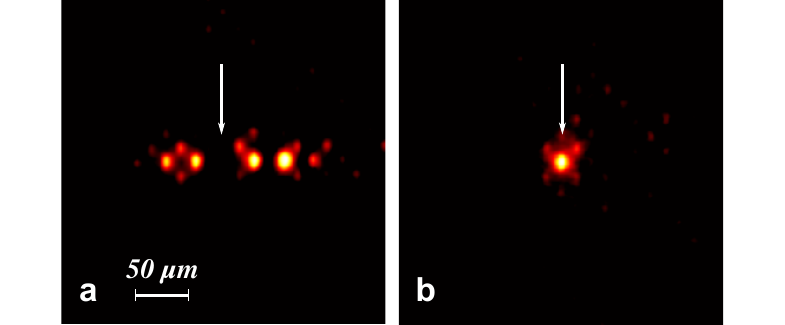}
\caption{{\bf Output light distribution at two propagation lengths}. Measured light distribution in the waveguide lattice section [corresponding to Fig.4(a)] at positions (a) $L_{max}$, where the fractional BO amplitude is at maximum and (b) $L_{foc}$ where light refocuses to the initial waveguide (indicated by a white arrow in both panels).}
\end{figure}
Figure 5(a)  clearly shows that, even in correspondence of the breath maximum amplitude, light remains confined in the waveguide lattice detuned diagonal. This fact is a further demonstration that two interacting particles, initially in the same lattice site, form a bound state and hop together to neighboring sites. Figure 5(b) shows that at the refocusing distance, all the light is indeed confined in the central waveguide, which was the one originally excited (indicated by a white arrow in both panels).\par

The 'fractional' nature of BO for correlated particles, namely frequency doubling with respect to the case of a single particle hopping in the same lattice under the action of the same force $F$, is shown in Fig.4(b). We fabricated a planar 1D array of $23$ identical waveguides, equally spaced of $d' = d =  19 \mu m$, and uniformly bent in the array plane with a constant radius of curvature $R'$. Such a structure reliably simulates the dynamics of single particle BO \cite{O5}. In order to implement the same force on the single particle we have to project the curvature on the diagonal for the two-particles structure on one of the main axis of the square lattice: this yields a curvature radius for the linear array $R'=R\sqrt{2}=400\sqrt{2}cm$. We imaged the light propagation in this waveguide array for a length $L'= L = 8.5 cm$ [Fig.4(b)]. Also in this case we observe a breathing propagation mode corresponding to the single particle BO, but the position of the maximum breath amplitude $L'_{max}$ corresponds to the refocusing position $L_{foc}$ for two-particles BO [Fig.4(a)]. This demonstrates that the frequency of a two-particles BO is twice that for a single-particle. It may be worth noting that the breathing amplitude of a single particle BO is larger than that for the two-particle BO. This is due to a higher hopping rate in the former case, $k>k_{eff}$, and to the double force experienced by the two strongly correlated particles with respect to the single one.\par

In conclusion, we have experimentally observed the photonic quantum-analogy of fractional Bloch Oscillations of two strongly correlated particles. We have shown that our photonic simulator is capable of implementing two-particle tunneling in addition to second-order pair tunneling, thus enabling the simulation of the extended Bose-Hubbard model. The simple visualization of the phenomenon under study and the high control on the fabricated structure, and thus of the implemented model, make this approach a very powerful tool to investigate other exotic phenomena associated to the formation of particle bound states  which are of difficult access in the matter.\par

{\small 
\acknowledgements{
This work was supported by the European Union through the project FP7-ICT-2011-9-600838 ({\it QWAD - Quantum Waveguides Application and Development}).}

\section*{Author contributions}
All authors conceived the experiment. G.C., A.C., R.O. designed and fabricated the photonic device and performed the measurements. G.D. and S.L. developed the theory underlying the experiment. All authors discussed the results and participated in the manuscript preparation.

\section*{Additional information}
Supplementary information is available in the online version of the paper.

\section*{Competing financial interests}
The authors declare no competing financial interests.

\section*{Methods}
{\bf Fabrication of the waveguide arrays.} Waveguide arrays have been fabricated by femtosecond laser waveguide writing in fused silica samples. The second harmonic of a HighQLaser femtoREGEN Yb-based amplified laser system has been used, consisting in 400-fs pulses at 520-nm wavelength with repetition rates up to 960 kHz. Actual writing conditions consisted in 300-nJ pulses delivered at a repetition rate of 20 kHz, reduced using the internal pulse-picker. Optimal range for the writing speed was identified as $8-14 mm/s$, yielding propagation losses of about $0.5 dB/cm$. In order to implement the detuning in the lattice diagonal we modulated the writing speed, therefore the waveguides on the diagonal were fabricated at a speed of $9 mm/s$, while all the others were written at $14 mm/s$. The variation in writing speed modifies the propagation constants of the guided modes, producing the designed detuning $\Delta \beta = - 4 cm^{-1}$, but has a negligible effect on the coupling rate between the waveguides. End-faces of the substrates were polished after writing to improve light launching and output imaging.\par

{\bf Characterization of the light distribution in the arrays.} Femtosecond laser writing in fused silica creates color centers that provide fluorescent emission at about 650 nm when light at 633 nm is propagated in the waveguide. Top-view imaging of the fluorescence signal is employed to visualize and quantitatively estimate the light distribution along the waveguide array, rejecting the
background light by a notch filter at 633 nm. In order to achieve a high resolution in imaging the light in the waveguide array all along its length, several images have been acquired and then stitched together. Propagation losses in the waveguides have been compensated by renormalizing the intensity levels in the acquired images.}

\bibliography{MEP_BIB}

\end{document}